\documentclass[%
reprint,superscriptaddress,
amsmath,amssymb,
aps,
]{revtex4-2}
\usepackage{hyperref}
\usepackage{graphicx}
\usepackage{dcolumn}
\usepackage{bm}
\usepackage{physics}
\usepackage{amsmath}
\usepackage{color}
\begin{document}
	\title{Sending-or-not-sending quantum key distribution with phase postselection}
	\author{Yang-Guang Shan}
	\author{Yao Zhou}
	\affiliation{CAS Key Laboratory of Quantum Information, University of Science and Technology of China, Hefei, Anhui 230026, China}
	\affiliation{CAS Center for Excellence in Quantum Information and Quantum Physics, University of Science and Technology of China, Hefei, Anhui 230026, China}
	\author{Zhen-Qiang Yin}
	\email{yinzq@ustc.edu.cn}
	\author{Shuang Wang}
	\email{wshuang@ustc.edu.cn}
	\author{Wei Chen}
	\author{De-Yong He}
	\author{Guang-Can Guo}
	\author{Zheng-Fu Han}
	\affiliation{CAS Key Laboratory of Quantum Information, University of Science and Technology of China, Hefei, Anhui 230026, China}
	\affiliation{CAS Center for Excellence in Quantum Information and Quantum Physics, University of Science and Technology of China, Hefei, Anhui 230026, China}
	\affiliation{Hefei National Laboratory, University of Science and Technology of China, Hefei 230088, China}
	
	\begin{abstract}
		Quantum key distribution (QKD) could help to share secure key between two distant peers. In recent years, twin-field (TF) QKD has been widely investigated because of its long transmission distance. One of the popular variants of TF QKD is sending-or-not-sending (SNS) QKD, which has been experimentally verified to realize 1000-km level fibre key distribution. In this article, the authors introduce phase postselection into the SNS protocol. With this modification, the probability of selecting "sending" can be substantially improved. The numerical simulation shows that the transmission distance can be improved both with and without the actively odd-parity pairing method. With discrete phase randomization, the variant can have both a larger key rate and a longer distance.
	\end{abstract}
	\maketitle
	\section{introduction}
	Quantum key distribution (QKD) \cite{bennett1984quantum} tries to establish secure communication between two distant peers, Alice and Bob, by sharing private random numbers. Guaranteed by the fundamental principles of quantum mechanics \cite{lo1999unconditional,shor2000simple}, Alice and Bob can  bound the amount of information stolen by any eavesdropper (usually called Eve), thus they could extract secure keys unknown to Eve.
	
	The basis of the security comes from the no-cloning theorem of a single photon \cite{wootters1982single}. Unluckily, because of the attenuation of the channels, a photon cannot transmit for a long distance. The PLOB bound \cite{pirandola2009direct,pirandola2017fundamental} gives the fundamental limitation of the point-to-point QKD, in which the key rate decreases linearly with the channel transmittance ($R\le O(\eta)$\cite{pirandola2017fundamental}) (see another bound in \cite{takeoka2014fundamental}). Fortunately, twin-field (TF) QKD \cite{lucamarini2018overcoming} broke this limitation reaching a key rate of $O(\sqrt{\eta})$ level. TF QKD has unique advantage for long-distance key distribution and has been widely researched both in theory\cite{ma2018phase,wang2018twin,lin2018simple,cui2019twin,curty2019simple} and experiment \cite{minder2019experimental,wang2019beating,liu2019experimental,zhong2019proof,fang2020implementation,chen2020sending,zhong2021proof,pittaluga2021600,clivati2022coherent,liu2021field,chen2021twin,wang2022twin,liu2023experimental}.
	
	Sending-or-not-sending (SNS) \cite{wang2018twin,jiang2019unconditional} QKD is a special kind of TF QKD, in which the information is encoded on the choice of sending a coherent state (sending) or a vacuum state (not sending). A distinctive advantage of SNS QKD is its high tolerance for misalignment error. Though the SNS protocol has a lot of advantages, there is a vital drawback. In SNS QKD, Alice and Bob randomly choose to send a coherent state with a probability $p$ or a vacuum state with a probability $1-p$. When they both choose to send the coherent state  and a successful click occurs, they get a bit error. Since the correct clicks correspond to the case that only one of Alice and Bob chooses to send a coherent state, whose counting rate is smaller than the sending-sending case, they must choose a small sending probability $p$ to reduce the probability of a sending-sending case ($p^2$). With a small $p$, the total counting rate is also reduced, limiting the performance of the protocol. 
	
	To solve this problem, the actively odd-parity pairing (AOPP) postprocessing method \cite{xu2020sending,jiang2020zigzag,jiang2021composable} can be used to significantly reduce the influence of bit errors and improve the sending probability. SNS-AOPP QKD can realize both a long transmission distance and a large key rate. Recently, utilizing the advantage of SNS-AOPP QKD, the first fibre-QKD over 1000 km is reported\cite{liu2023experimental}. 
	
	In this article, we give another way to reduce the bit error rate by introducing phase postselection into SNS QKD. In the original SNS protocol, the phases of the coherent states are randomized. But when the phases of Alice and Bob are the same in the sending-sending cases, more left-click events (constructive interference) will happen than right-click events (destructive interference) in an interferometer. And if the phase difference between Alice and Bob is $\pi$, more right-click events will happen. So before the postprocessing of the keys, Alice and Bob can announce their phases of all rounds. If the phases of Alice and Bob are close, they only keep the right-click events. And if the phases are opposite, they only keep the left-click events. This sifting step may discard a part of the correct bits but almost all error bits. Thus a large sending probability $p$ can be chosen and the total number of correct bits can be improved.
	
	Naturally, the behavior of announcing the phases will increase the information that an eavesdropper can get. The coherent state sent by Alice or Bob cannot be treated as a mixed state of Fock states. Thus we give the security analysis of the new protocol in this article and compare the performance with the original SNS protocol. 
	
	We analyzed several kinds of variants of our modification. The phases of signal states can be continuously randomized or discretely randomized. We also try to apply the AOPP method to improve the transmission distance. Our numerical simulation shows that with continuous phase randomization, the transmission distance can be much longer than the original SNS protocol. With discrete phase randomization, our variant can have both a larger key rate and a longer transmission distance. The AOPP method could drastically improve the performance of the original SNS protocol. However, the improvement is not so large if AOPP is used in our protocol. AOPP cannot help to improve the key rate of our variant, but the transmission distance is improved distinctly. Both with AOPP, our variant could have a longer distance than the SNS-AOPP protocol.
	
	In Section.(\ref{protocol}) we introduce the procedure of our protocol. In Section.(\ref{secu}) we give the security analysis of our protocol. In Section.(\ref{sim}) we conduct the numerical simulation to compare the performance of our protocol. In Section.(\ref{dis}) we discuss some variants of our protocol. We come to conclusion in Section.(\ref{conclude}).
\section{protocol description\label{protocol}}
We introduce the procedure of our variant with continuous phase randomization in the following. For ease of understanding, we only give the flow of signal states.
\begin{enumerate}
	\item 
	\textbf{State preparation.} Alice (Bob) randomly chooses to send a weak coherent state with a probability $p$ or to send a vacuum state with a probability $1-p$. When Alice (Bob) chooses to send a weak coherent state, she (he) records a local classical bit 1 (0), else she (he) records a local classical bit 0 (1).

	 When Alice (Bob) decides to send a weak coherent state, she (he) prepares a phase-randomized weak coherent state $\ket{\alpha \text{e}^{i\theta_A}}$ ($\ket{\alpha \text{e}^{i\theta_B}}$) and saves the phase $\theta_A$ ($\theta_B$) locally. 
	 
	 When Alice (Bob) decides to send a vacuum state $\ket{0}$, she (he) produces a uniform random phase in $[0,2\pi)$ and saves it as the phase $\theta_A$ ($\theta_B$).
	 
	 Alice and Bob send the states to the third party Charlie in the middle of the channel.
	 
	 In this protocol, we separate all the rounds into two kinds. In $\mathbb{C}$ (correct) rounds, Alice and Bob record the same classical bits, which means only one of them selects to send the vacuum state and the other chooses to send the coherent state. In $\mathbb{E}$ (error) rounds, Alice and Bob record different classical bits, which means Alice and Bob both select to send the vacuum states or both select to send the coherent states.
	 
	 \item 
	 \textbf{State measurement.} If Charlie is honest, he will perform an interferometric measurement with the two states from Alice and Bob. We assume that the left detector corresponds to the constructive interference and the right detector corresponds to the destructive interference for pulses with the same phase. Then Charlie will declare a left-click event, a right-click event or a failed event according to the clicks of the two detectors. Left-click events and right-click events are collectively called successful events.

	 \item
	 \textbf{Phase postselection.} After enough rounds of the first two steps, Alice and Bob keep the rounds with successful events and discard the others. Then Alice and Bob announce the phases they saved in the first step publicly. For every left-click event, if $\abs{\abs{\theta_A-\theta_B}-\pi}\le\Delta$, they keep the round as a sifted left-click event. And for every right-click event, if $\abs{\theta_A-\theta_B}\le\Delta$ or $\abs{\abs{\theta_A-\theta_B}-2\pi}\le\Delta$, they keep the round as a sifted right-click event.
	 
	 \item 
	 \textbf{Parameter estimation and postprocessing.} Alice and Bob use decoy states \cite{hwang2003quantum,PhysRevLett.94.230504,wang2005beating} to estimate the phase error rates. Then Alice and Bob conduct error correction and privacy amplification to the classical bits of the remaining rounds to get the final key. They may use some two-way error rejection methods to improve the performance of the protocol, for example, the AOPP method.

\end{enumerate}
\section{Security analysis\label{secu}}

Notice that there are two kinds of successful events, the $\mathbb{C}$-round events and the $\mathbb{E}$-round events. We cannot separate them before the error correction, thus they both contribute to the key consumption of error correction. Similar with the analysis of the original SNS protocol, we only use the $\mathbb{C}$ rounds to generate keys.

Firstly, we give the equivalent protocol based on entanglement. In the following, we will analyze the security of right-click events, and the security of left-click events is analogous.

Because of the phase interval $\Delta$ of the phase postselection, the phase difference between Alice and Bob is set to $\delta\in[-\Delta,\Delta]$. Then we can give the equivalent protocol, in which Alice and Bob prepare the state,
\begin{equation}
	\begin{aligned}
		\ket{\psi(\theta,\delta)}=(\sqrt{1-p}\ket{0}_A\ket{0}_a+\sqrt{p}\ket{1}_A\ket{\alpha \text{e}^{i\theta}}_a)\\
		\otimes(\sqrt{1-p}\ket{1}_B\ket{0}_b+\sqrt{p}\ket{0}_B\ket{\alpha\text{e}^{i(\theta+\delta)}}_b),
	\end{aligned}\label{entanglesame}
\end{equation} 
where the subscripts $A$ and $B$ correspond to the local ancillas of Alice and Bob, and the subscripts $a$ and $b$ correspond to the states sent to Charlie. To get the classical bits, Alice and Bob can measure their ancillas on the $\mathbb{Z}$ basis ($\ket{0}$ and $\ket{1}$). And the phase error rate, which is the error rate when Alice and Bob measure their ancillas on the $\mathbb{X}$ basis ($\ket{+}=(\ket{0}+\ket{1})/\sqrt{2}$ and $\ket{-}=(\ket{0}-\ket{1})/\sqrt{2}$), relates to the amount of information that may be leaked to eavesdroppers. $\theta$ is the phase of Alice, which is randomized in all rounds. $\delta$ is also randomized from $[-\Delta,\Delta]$.

For the $\mathbb{C}$ rounds, only one of Alice and Bob decides to send a weak coherent state. Thus a state in a $\mathbb{C}$ round can be written as 
\begin{equation}
	\begin{aligned}
	&(\ket{00}_{AB}\bra{00}_{AB}+\ket{11}_{AB}\bra{11}_{AB})\ket{\psi(\theta,\delta)}\\=&\sqrt{p(1-p)}(\ket{00}_{AB}\ket{0}_a\ket{\alpha\text{e}^{i(\theta+\delta)}}_b+\ket{11}_{AB}\ket{\alpha\text{e}^{i\theta}}_a\ket{0}_b).
	\end{aligned}\label{keygene}
\end{equation}

We define that when Alice and Bob measure their ancillas on the $\mathbb{X}$ basis, $\ket{+-}_{AB}$ and $\ket{-+}_{AB}$ correspond to correct results, and $\ket{++}_{AB}$ and $\ket{--}_{AB}$ correspond to errors. Operating $\ket{++}\bra{++}_{AB}+\ket{--}\bra{--}_{AB}$ on eq.(\ref{keygene}), the probability of a $\mathbb{C}$-round right-click phase error, when a round of $\ket{\psi(\theta,\delta)}$ is sent, can be estimated as $P_{ph}^R(\theta,\delta)=p(1-p)P^R(\frac{\ket{0}_a\ket{\alpha\text{e}^{i(\theta+\delta)}}_b+\ket{\alpha\text{e}^{i\theta}}_a\ket{0}_b}{\sqrt{2}})$. Here $P^R(\ket{\cdot})$ means the probability that a right-click event is declared by Charlie when a state $\ket{\cdot}$ is sent from Alice and Bob. $\ket{\cdot}$ can be unnormalized, and $P^R(c\ket{\cdot})=\abs{c}^2P^R(\ket{\cdot})$.

Realizing that the phase $\theta$ is randomly chosen in $[0,2\pi)$ and $\delta$ is also randomized, to get the average of $P_{ph}^R$, we calculate the integration below in eq.(\ref{pph}). Here $\ket{j}$ is the Fock state of $j$ photons. $\mu=\abs{\alpha}^2$ is the intensity of the coherent state.
\begin{widetext}

\begin{equation}
	\begin{aligned}
		P^R_{ph}=&\frac{1}{4\pi\Delta}\int_{-\Delta}^{\Delta}d\delta\int_0^{2\pi}d\theta P_{ph}^R(\theta,\delta)\\
		=&p(1-p)\text{e}^{-\mu}\left(2P^R(\ket{0}_a\ket{0}_b)+\sum_{j=1}^\infty\frac{\mu^j}{j!} \frac{1}{2\Delta}\int_{-\Delta}^{\Delta}d\delta P^R(\frac{\text{e}^{ij\delta}\ket{0}_a\ket{j}_b+\ket{j}_a\ket{0}_b}{\sqrt{2}})\right)
	\end{aligned}\label{pph}
\end{equation}
\end{widetext}

In eq.(\ref{pph}) we use the linear additivity between the measurement probability and the density matrix. The probability of a right-click event can be calculated with a measurement matrix $M^R$. Then eq.(\ref{pph}) can be given by eq.(\ref{tr}), where $\mathcal{P}(\ket{\cdot})=\ket{\cdot}\bra{\cdot}$.
\begin{widetext}
\begin{equation}
	\begin{aligned}
		\frac{1}{2\pi}\int_0^{2\pi}d\theta P^R(\frac{\ket{0}_a\ket{\alpha\text{e}^{i(\theta+\delta)}}_b+\ket{\alpha\text{e}^{i\theta}}_a\ket{0}_b}{\sqrt{2}})=&\frac{1}{2\pi}\int_0^{2\pi}d\theta \text{Tr} \left(M^R\mathcal{P}(\frac{\ket{0}_a\ket{\alpha\text{e}^{i(\theta+\delta)}}_b+\ket{\alpha\text{e}^{i\theta}}_a\ket{0}_b}{\sqrt{2}})\right)\\
		=&\text{Tr}\left(M^R\frac{1}{2\pi}\int_0^{2\pi}d\theta\mathcal{P}(\sum_{j=0}^\infty\sqrt{\frac{\text{e}^{-\mu}\mu^j}{j!}}\text{e}^{ij\theta}\frac{\text{e}^{ij\delta}\ket{0}_a\ket{j}_b+\ket{j}_a\ket{0}_b}{\sqrt{2}})\right)\\
		=&\text{Tr}\left(M^R\left(\text{e}^{-\mu}\mathcal{P}(\sqrt{2}\ket{00}_{ab})+\sum_{j=1}^\infty\text{e}^{-\mu}\frac{\mu^j}{j!}\mathcal{P}(\frac{\text{e}^{ij\delta}\ket{0}_a\ket{j}_b+\ket{j}_a\ket{0}_b}{\sqrt{2}})\right)\right)\\
		=&2\text{e}^{-\mu}\Tr(M^R\mathcal{P}(\ket{00}_{ab}))+\sum_{j=1}^\infty\text{e}^{-\mu}\frac{\mu^j}{j!}\Tr(M^R\mathcal{P}(\frac{\text{e}^{ij\delta}\ket{0}_a\ket{j}_b+\ket{j}_a\ket{0}_b}{\sqrt{2}})).
	\end{aligned}\label{tr}
\end{equation}
\end{widetext}

The average right-click rate of states $\text{e}^{ij\delta}\ket{0}_a\ket{j}_b+\ket{j}_a\ket{0}_b$ can be estimated by decoy states, which is explained in detail in Appendix.\ref{est}.

The phase error rate of $\mathbb{C}$ rounds can be given by $e_{ph}^R=\frac{P_{ph}^R}{P_c^R}$, where $P_c^R$ is the probability that a $\mathbb{C}$-round right-click event occurs when $\ket{\psi}$ is sent. Then the key rate of right-click events is shown as 
\begin{equation}
	R^R=s (P_c^R(1-H_2(\frac{P_{ph}^R}{P_c^R}))-f P_t^R H_2(e_{bit}^R)),\label{rate}
\end{equation}
where $H_2(x)=-x\log_2(x)-(1-x)\log_2(1-x)$ is the binary entropy function, and $f$ is the error correction efficiency. $s=\frac{\Delta}{\pi}$ is the phase sifting efficiency, and $P_t^R$ is the right-click rate of sifted rounds. $e_{bit}^R$ is the bit error rate of sifted right-click rounds. $P_t^R$ is known by counting the number of sifted right-click events and $P_c^R=P_t^R(1-e_{bit})$ is simply known after the error correction step. 

For the left-click events, the analysis is similar. The only difference is that the phase error corresponds to the left-click rate of $\frac{\ket{0}_a\ket{\alpha\text{e}^{i(\theta+\delta)}}_b+\ket{-\alpha\text{e}^{i\theta}}_a\ket{0}_b}{\sqrt{2}}$. The key rate is shown as
\begin{equation}
	R^L=s (P_c^L(1-H_2(\frac{P_{ph}^L}{P_c^L}))-f P_t^L H_2(e_{bit}^L)).
\end{equation}
And the total key rate can be shown as $R=R^R+R^L$.

\section{numerical simulation\label{sim}}
We conduct numerical simulation to show the advantage of our modification. In our simulation, infinite decoy states are assumed. Thus we can directly calculate the estimation of phase errors. The detailed calculation is shown in Appendix.\ref{dsim}. The device parameters of the simulation are shown in Table. (\ref{table:pa}), where $P_d$ is the detecting efficiency of detectors. $d$ is the dark counting rate of detectors. $f$ is the error correction efficiency. $e_{\text{mis}}$ is the misalignment error rate.  

\begin{table}[h]
	\caption{\label{table:pa}Parameters we used in our simulation.}
	\begin{ruledtabular}
		\begin{tabular}{cccc}
			$P_d$&$d$&$f$&$e_\text{mis}$\\\hline
			$1$&$10^{-11}$&$1.1$&$0.01$
		\end{tabular}
	\end{ruledtabular}
\end{table}

\begin{figure}[h]
	\includegraphics[width=0.5\textwidth]{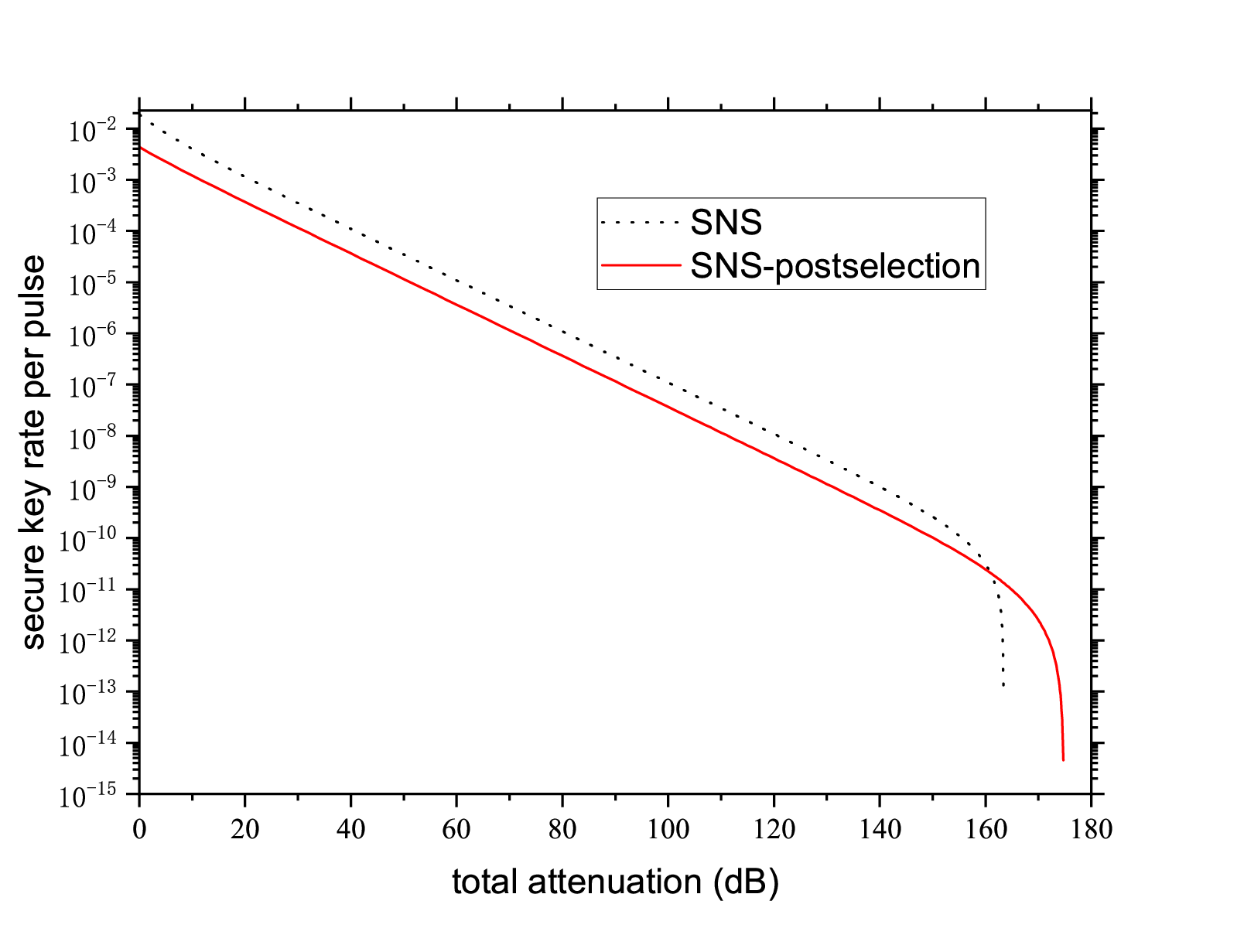}
	\caption{\label{fig:sns-lock}The simulation result of the original SNS protocol and our modified protocol. The line SNS-postselection corresponds to the key rates of our variant, and the line SNS corresponds to the key rate of the original SNS protocol.}
\end{figure}

We compare the key rates of the original SNS protocol and our variant, which is shown in Fig.(\ref{fig:sns-lock}). We can see that our variant has a longer transmission distance of additional 10 dB, which corresponds to about a 50-km fibre channel. 

In the original SNS protocol, the optimal sending probability is about $5\%$, which means that in $90\%$ of all rounds both Alice and Bob send vacuum states. Thus at long distance, the dark counts of these rounds increase the bit errors a lot. The $\mathbb{C}$ rounds only account for about $10\%$. Thus the signal to noise ratio is significantly influenced at long distance. And in our variant, the optimal sending probability increases with the distance and reaches more than $40\%$ at extreme distance, which means sending-sending rounds and vacuum-vacuum rounds account for about $52\%$. The rest $48\%$ rounds are $\mathbb{C}$ rounds. Thus a longer distance is reasonable.
\section{discussion\label{dis}}
In our numerical simulation, we have compared the performance of our modified protocol and the original SNS protocol. It seems that our protocol could reach a longer transmission distance, but the key rate is relatively lower. For the case of short transmission distance, the sifting efficiency must be the major restriction of the key rates. A small sifting interval $\Delta$ will decrease the sifting efficiency, while a large sifting interval may increase the bit error rate and the phase error rate. To beat the SNS protocol at short distance, we give another variant of our protocol below.

In step. 1 of the protocol description, when Alice (Bob) chooses to send a weak coherent state, she (he) will randomly choose a phase $\theta_A$ ($\theta_B$) from $\{0,\frac{2\pi}{M},2\frac{2\pi}{M},\dots,(M-1)\frac{2\pi}{M}\}$ ($M>0$ is an even number) and prepare a state $\ket{\alpha \text{e}^{i\theta_A}}$ ($\ket{\alpha \text{e}^{i\theta_B}}$). When she (he) chooses to send a vacuum state, she (he) also records a random phase $\theta_A$ ($\theta_B$) from $\{0,\frac{2\pi}{M},2\frac{2\pi}{M},\dots,(M-1)\frac{2\pi}{M}\}$. In step. 3, for left-click events, the rounds that $\abs{\theta_A-\theta_B}=\pi$ are kept. And for right-click events, the rounds that $\theta_A=\theta_B$ are kept. Thus the sifting efficiency is about $1/M$. $M$ is set to be an even number to keep the existence of $\abs{\theta_A-\theta_B}=\pi$ events. Note that $M=1$ is also feasible, and only one detector is needed (the right detector) to conduct the protocol. But we do not discuss it because of its low performance. 

\begin{figure}[h]
	\centering
	\includegraphics[width=0.5\textwidth]{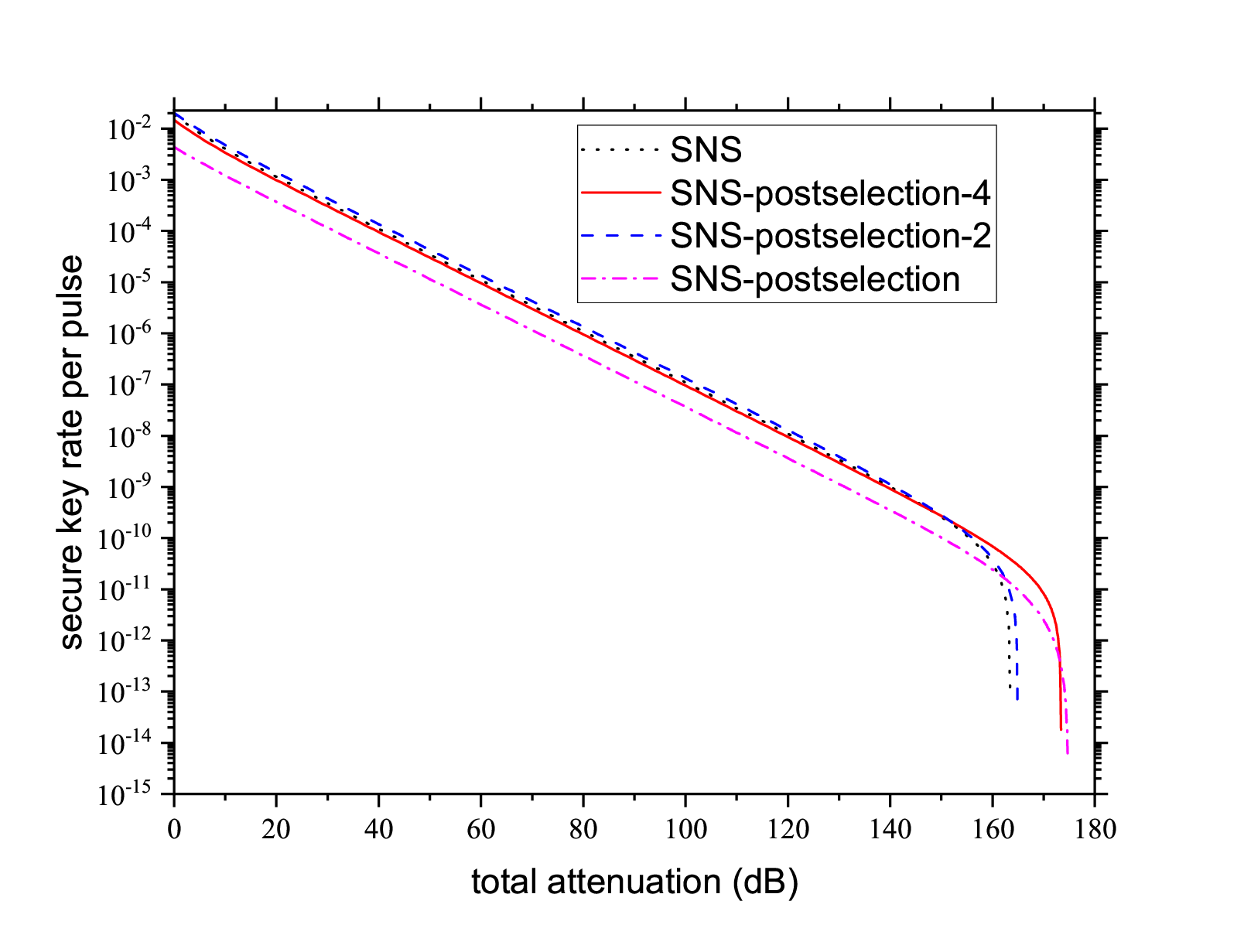}
	\caption{\label{fig:24}The simulation result of our variants with discrete phase randomization. The line SNS-postselection-2 corresponds to the case of two random phases ($M=2$). The line SNS-postselection-4 corresponds to the case of four random phases ($M=4$). }
\end{figure}

$M=2,4$ might be two good choices, since a big $M$ also decreases the sifting efficiency. We give the security analysis of the $M=2,4$ cases in the Appendix.\ref{discrete}. We also simulated the key rates of these two cases shown in Fig.(\ref{fig:24}).

The simulation shows that with discrete phase randomization, the key rate of our variant is improved, which can approach or exceed the key rate of the original SNS protocol. The $M=2$ case can both have a larger key rate and a longer distance than the original SNS protocol. The $M=4$ case can approach the maximum distance of the continuous-phase-randomization case and have a larger key rate. 

\begin{figure}
	\centering
	\includegraphics[width=0.5\textwidth]{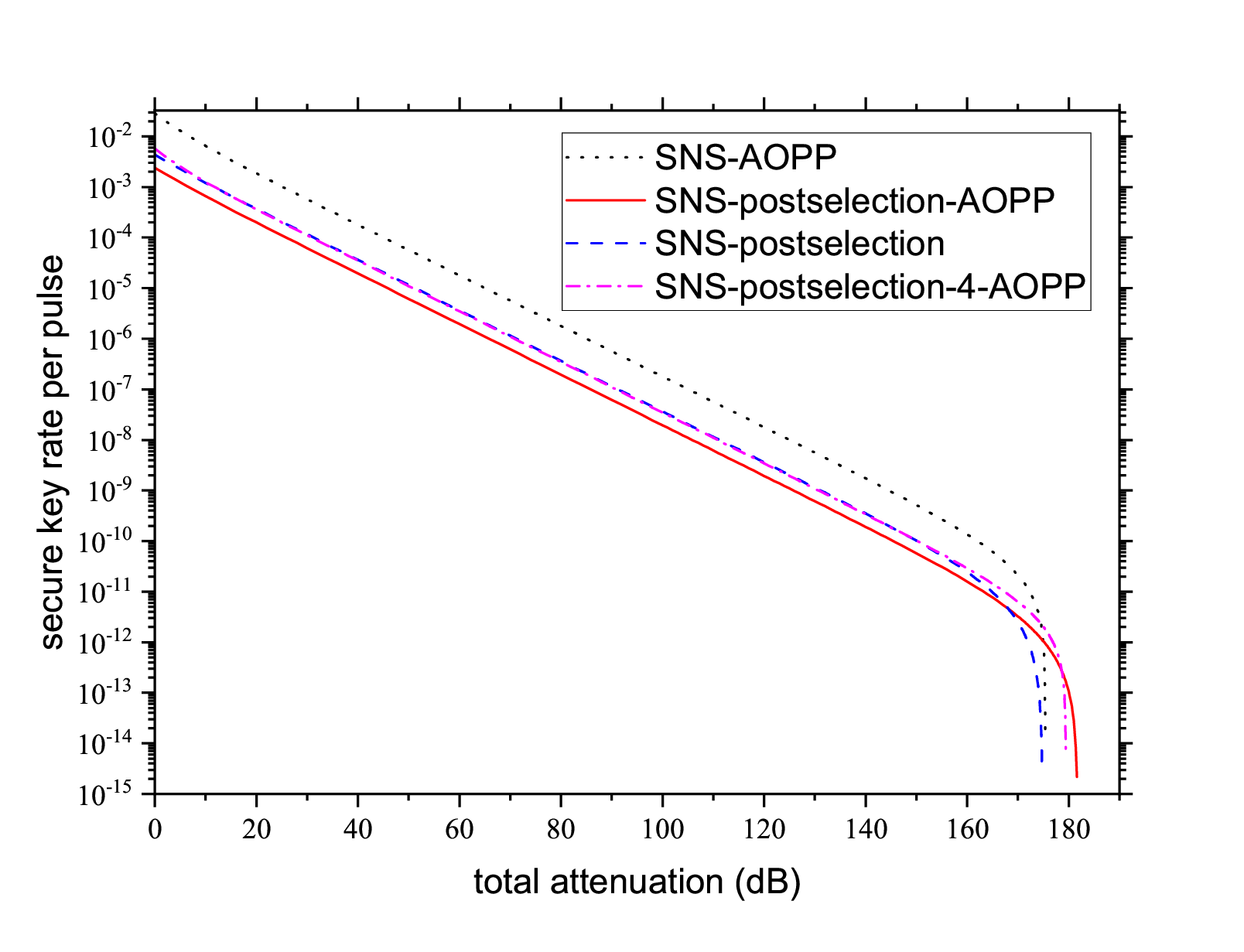}
	\caption{\label{fig:aopp}The simulation result of SNS protocol and our variant with AOPP. The line SNS-AOPP corresponds to the case of the original SNS with AOPP postprocessing. The line SNS-postselection-AOPP corresponds to the case of our continuous-phase-randomization variant with AOPP. The line SNS-postselection-4-AOPP corresponds to our four-phase variant with AOPP.}
\end{figure}

The original SNS protocol benefits a lot from the AOPP postprocessing method, which both improves the key rate and the transmission distance of the SNS protocol. The SNS-AOPP protocol has been widely used to realize long-distance QKD \cite{liu2021field,chen2021twin,liu2023experimental}. Realizing that the $\mathbb{C}$ rounds of our variant have no bit errors, we can directly use the security analysis of the SNS-AOPP protocol \cite{xu2020sending} to apply AOPP on our variant. We compare the performance with AOPP in Fig.(\ref{fig:aopp}). The simulation shows that SNS-AOPP has an obvious advantage on key rate, but its transmission distance is the same as our variant without AOPP. AOPP does not help to improve the key rate of our variant. However, with AOPP our variant could have a longer transmission distance overwhelming the SNS-AOPP.

\section{Conclusion\label{conclude}}
To conclude, we proposed a new kind of variant of the SNS QKD by introducing phase postselection. We also analyzed the performance of our variant with discrete phase randomization and AOPP two-way postprocessing. We gave the security analysis of our variant and performed numerical simulations to compare these protocols. 

Without two-way postprocessing, the continuous-phase-randomization case and the four-phase discrete-phase-randomization case of our variant can reach a much longer distance than the original SNS protocol. The two-phase discrete-phase-randomization case of our variant can have a both larger key rate and a longer distance than the SNS protocol.

With AOPP two-way postprocessing, the SNS-AOPP protocol can have a larger key rate than our variant. However, our variant still has an advantage in transmission distance.

The additional step of phase sifting in our variants does not increase the difficulty of experimental realization, because a similar phase announcement is needed for the decoy states in the original SNS protocol. To realize QKD at an extremely long distance, our variant with AOPP might be a better choice than the SNS-AOPP protocol. 
\begin{acknowledgments}
This work has been supported by the National Key Research and Development Program of China (Grant No. 2020YFA0309802), the National Natural Science Foundation of China (Grant Nos. 62171424 and 62271463), Prospect and Key Core Technology Projects of Jiangsu provincial key R \& D Program (BE2022071), the Fundamental Research Funds for the Central Universities, the Innovation Program for Quantum Science and Technology (Grant No. 2021ZD0300701).
\end{acknowledgments}

\appendix
\section{Feasibility of the decoy state estimation\label{est}}
The given probability of a phase error from the main text is shown as 
\begin{equation}
	\begin{aligned}
		&P^R_{ph}
		=p(1-p)\text{e}^{-\mu}\Bigg(2P^R(\ket{0}_a\ket{0}_b)\\
		&+\sum_{j=1}^\infty\frac{\mu^j}{j!} \frac{1}{2\Delta}\int_{-\Delta}^{\Delta}d\delta P^R(\frac{\text{e}^{ij\delta}\ket{0}_a\ket{j}_b+\ket{j}_a\ket{0}_b}{\sqrt{2}})\Bigg).
	\end{aligned}
\end{equation}
Similarly, the phase error rate of left-click events is 
\begin{equation}
	\begin{aligned}
		&P^L_{ph}
		=p(1-p)\text{e}^{-\mu}\Bigg(2P^L(\ket{0}_a\ket{0}_b)\\
		&+\sum_{j=1}^\infty\frac{\mu^j}{j!} \frac{1}{2\Delta}\int_{-\Delta}^{\Delta}d\delta P^L(\frac{\text{e}^{ij\delta}\ket{0}_a\ket{j}_b+(-1)^j\ket{j}_a\ket{0}_b}{\sqrt{2}})\Bigg).
	\end{aligned}
\end{equation}

To estimate the phase error rate, we need to know the average right-click rates of the states $\frac{\text{e}^{ij\delta}\ket{0}_a\ket{j}_b+\ket{j}_a\ket{0}_b}{\sqrt{2}}$ and the average left-click rate of the states $\frac{\text{e}^{ij\delta}\ket{0}_a\ket{j}_b+(-1)^j\ket{j}_a\ket{0}_b}{\sqrt{2}}$. Luckily, the decoy state method can help us to complete this estimation.

Firstly the click rate of $\ket{00}$ can be easily estimated with vacuum decoy state. 

Then with phase locking and phase postselection, we also keep the events that Alice and Bob send decoy states with the same intensity $\nu$ and the phase difference less than $\Delta$. Thus the density matrix is shown as
\begin{equation}
	\small
	\begin{aligned}
	\frac{1}{4\pi\Delta}\int_{-\Delta}^{\Delta}d\delta\int_0^{2\pi}d\theta \ket{\sqrt{\nu}\text{e}^{i\theta}}\ket{\sqrt{\nu}\text{e}^{i(\theta+\delta)}}\bra{\sqrt{\nu}\text{e}^{i\theta}}\bra{\sqrt{\nu}\text{e}^{i(\theta+\delta)}}\\
	=\text{e}^{-2\nu}\ket{00}\bra{00}+\text{e}^{-2\nu}2\nu\frac{1}{2\Delta}\int_{-\Delta}^{\Delta}d\delta\mathcal{P}(\frac{\text{e}^{i\delta}\ket{01}+\ket{10}}{\sqrt{2}})\\
	+\text{e}^{-2\nu}\frac{(2\nu)^2}{2}\frac{1}{2\Delta}\int_{-\Delta}^{\Delta}d\delta\mathcal{P}(\frac{\text{e}^{2i\delta}\ket{02}+\ket{20}+\sqrt{2}\text{e}^{i\delta}\ket{11}}{2})+\dots
\end{aligned}\label{decoysame}
\end{equation}
Then the right-click rate of this state is shown as 
\begin{equation}
	\small
	\begin{aligned}
	&\text{e}^{-2\nu}P^R(\ket{00})+\text{e}^{-2\nu}2\nu \left( \bar P^R(\frac{\text{e}^{i\delta}\ket{01}+\ket{10}}{\sqrt{2}})\right)\\
	&+\text{e}^{-2\nu}\frac{(2\nu)^2}{2}\left(\bar P^R(\frac{\text{e}^{2i\delta}\ket{02}+\ket{20}+\sqrt{2}\text{e}^{i\delta}\ket{11}}{2})\right)+\dots
	\end{aligned}
\end{equation}
Here $\bar P^R$ is the average right-click rate for $\delta\in[-\Delta,\Delta]$. The click rate of this state can be expressed as a linear combination of states $\ket{00}$, the average of states $\frac{\text{e}^{i\delta}\ket{01}+\ket{10}}{\sqrt{2}}$ and so on. Thus known the click rates of the states with different $\nu$, we can get the click rates of $\ket{00}$, the average click rate of states $\frac{\text{e}^{i\delta}\ket{01}+\ket{10}}{\sqrt{2}}$ and the average click rate of states $\frac{\text{e}^{2i\delta}\ket{02}+\ket{20}+\sqrt{2}\text{e}^{i\delta}\ket{11}}{2}$ with linear programming. With a same reason, using the events that Alice and Bob have opposite phases, we can get the average click rate of $\frac{\text{e}^{2i\delta}\ket{02}+\ket{20}-\sqrt{2}\text{e}^{i\delta}\ket{11}}{2}$. This kind of estimation is practical and has been used in phase-matching QKD \cite{ma2018phase}.

With the decoy states that the phases of Alice and Bob are randomized separately, Alice and Bob can estimate the click rates of states $\ket{11}$, $\ket{0j}$ and $\ket{j0}$ ($j=3,4,\dots$) with linear programming. Now, knowing the average click rates of $\frac{\text{e}^{2i\delta}\ket{02}+\ket{20}+\sqrt{2}\text{e}^{i\delta}\ket{11}}{2}$, $\frac{\text{e}^{2i\delta}\ket{02}+\ket{20}-\sqrt{2}\text{e}^{i\delta}\ket{11}}{2}$ and $\ket{11}$ we can get the click rate of $\frac{\text{e}^{2i\delta}\ket{02}+\ket{20}}{\sqrt{2}}$ because of the following equation.

\begin{equation}\begin{aligned}
	&\frac{1}{2\Delta}\int_{-\Delta}^{\Delta}d\delta\Bigg(\mathcal{P}(\frac{\text{e}^{2i\delta}\ket{02}+\ket{20}+\sqrt{2}\text{e}^{i\delta}\ket{11}}{2})\\
	&+\mathcal{P}(\frac{\text{e}^{2i\delta}\ket{02}+\ket{20}-\sqrt{2}\text{e}^{i\delta}\ket{11}}{2})\Bigg)\\
	=&\mathcal{P}(\ket{11})+\frac{1}{2\Delta}\int_{-\Delta}^{\Delta}d\delta\mathcal{P}(\frac{\text{e}^{2i\delta}\ket{02}+\ket{20}}{\sqrt{2}}).\end{aligned}
\end{equation}
With the counting rates of $\ket{0j}$ and $\ket{j0}$, we can get the upper bound of the click rate of $\frac{\text{e}^{ij\delta}\ket{0j}+\ket{j0}}{\sqrt{2}}$ because of the following equation.
\begin{equation}
	\begin{aligned}
	\mathcal{P}(\frac{\text{e}^{ij\delta}\ket{0j}+\ket{j0}}{\sqrt{2}})+\mathcal{P}(\frac{\text{e}^{ij\delta}\ket{0j}-\ket{j0}}{\sqrt{2}})=\mathcal{P}(\ket{0j})+\mathcal{P}(\ket{j0})
	\end{aligned}
\end{equation}
\begin{equation}
	P^R(\frac{\text{e}^{ij\delta}\ket{0j}+\ket{j0}}{\sqrt{2}})\le P^R(\ket{0j})+P^R(\ket{j0})
\end{equation}
Without an eavesdropper, the upper bound is about the twice of the realistic click rate. However, we only use this upper bound for $j\ge3$, which do not influence a lot of the key rate.

The estimation of the left-click rate is the same. Yet we have finish the estimation of the phase error.

\section{Details of our simulation\label{dsim}}
In this section, we will introduce the calculation of our simulation. In the simulation, it is reasonable to assume the key rates of the two detectors are the same. Thus $R=R^R+R^L=2R^R$.

$P_c^R$ is the probability that a right-click $\mathbb{C}$-round event happens when the phases pass the sifting. It can be calculated as 
\begin{equation}
	P_c^R=2p(1-p)(1-\text{e}^{-\sqrt{\eta}\mu/2}(1-d)).
\end{equation} 
Here $2p(1-p)$ is the probability of a $\mathbb{C}$ round. Alice's coherent state $\ket{\alpha}$ of $\ket{\alpha}_a\ket{0}_b$ passes half of the channel with a transmittance $\sqrt{\eta}$ and then passes a beam splitter of Charlie to reach the right detector. We do not care the left detector here, thus the beam splitter can be treated as a attenuation with a transmittance of $\frac{1}{2}$. 

With infinite decoy states, we directly calculate every items of $P_{ph}^R$ which are shown in eq.(\ref{itemph}). For the single-photon item, with a phase difference of $\delta$, the probability that the photon enters the right detector becomes $\frac{1}{2}(1-\cos(\delta))$. For the two-photon item, when $\delta=0$ the two photons will enter a same detector simultaneously, while $\delta\ne0$ causes that sometimes the two photons transmit independently. We use the worst case of independent transmission to calculate the click rate. For the items with three or more photons, we use the upper bound shown in Appendix.\ref{est} to calculate the twice of the click rate of the state $\ket{j0}_{ab}$. Here $e_{\text{mis}}$ has no influence for the items with two or more photons.
\begin{widetext}
\begin{equation}
	\begin{aligned}
	P^R(\ket{00}_{ab})=&d\\
	P^R(\frac{\text{e}^{i\delta}\ket{01}_{ab}+\ket{10}_{ab}}{\sqrt{2}})=&\sqrt{\eta}(1-\left(\frac{1}{2}(1+\cos(\delta))(1-e_{\text{mis}})+\frac{1}{2}(1-\cos(\delta))e_{\text{mis}}\right)(1-d))+(1-\sqrt{\eta})d\\
	P^R(\frac{\text{e}^{2i\delta}\ket{02}_{ab}+\ket{20}_{ab}}{\sqrt{2}})=&(1-\sqrt{\eta}/2)^2 d+1-(1-\sqrt{\eta}/2)^2\\
	P^R(\frac{\text{e}^{ij\delta}\ket{0j}_{ab}+\ket{j0}_{ab}}{\sqrt{2}})\le&2((1-\sqrt{\eta}/2)^j d+1-(1-\sqrt{\eta}/2)^j)\\
	&(j\ge3)
	\end{aligned}\label{itemph}
\end{equation}
\end{widetext}

To get the $P_t^R$, we must use integration to get the click rate when Alice and Bob both choose to send a coherent state. $P_t^R$ is shown in eq.(\ref{pt}), and the bit error rate is $e_{bit}=P_E^R/P_t^R$.
\begin{widetext}
	\begin{equation}\begin{aligned}
	P_{E}^R=&(1-p)^2d+p^2\int_{-\Delta}^{\Delta}d\delta(1-(1-d)\text{e}^{-\sqrt{\eta}\mu(1-(1-2e_{\text{mis}})\cos(\delta))})/(2\Delta )\\
	P_{t}^R=&P_c^R+P_E^R
	\end{aligned}\label{pt}
	\end{equation}
\end{widetext}

With above equations, the key rate of our continuous-phase-randomization can be calculated. 
\section{Security of discrete phase randomization\label{discrete}}
To increase the key rate by improving the sifting efficiency, we consider the case that Alice and Bob only prepare their states of two or four phases. The protocol modification has been introduced in the main body of the article in Section.\ref{dis}. In the following we give the security analysis of the cases that $M=2,4$ separately.
\subsection{$M=2$}
In the case of $M=2$, Alice and Bob only prepare their states in two phases $\{0,\pi\}$. The sifting efficiency is $1/2$.

For right-click events, the sifting condition is $\theta_A=\theta_B\equiv\theta$. At the condition of a fixed common phase, the equivalent protocol and the phase error probability is the same as the one in Section.(\ref{secu}) with $\delta=0$. The state prepared by Alice and Bob can be set to eq.(\ref{entanglesame}) with $\delta=0$, and the phase error probability is $P_{ph}^R(\theta,0)=p(1-p)P^R(\frac{\ket{0}_a\ket{\alpha\text{e}^{i\theta}}_b+\ket{\alpha\text{e}^{i\theta}}_a\ket{0}_b}{\sqrt{2}})$. The difference is that the average phase error probability $P_{ph}^{R-2}$ is for $\theta\in\{0,\pi\}$.
\begin{widetext}
\begin{equation}
	\begin{aligned}
	P_{ph}^{R-2}=\frac{1}{2}\sum_{\theta=0,\pi}P_{ph}^R(\theta,0)=p(1-p)\Big\{ P^R\left(\sum_{j=0}^{\infty}\sqrt{\frac{\text{e}^{-\mu}\mu^{2j}}{(2j)!}}\frac{\ket{0}_a\ket{2j}_b+\ket{2j}_a\ket{0}_b}{\sqrt{2}}\right)\\
	+P^R\left(\sum_{j=0}^{\infty}\sqrt{\frac{\text{e}^{-\mu}\mu^{2j+1}}{(2j+1)!}}\frac{\ket{0}_a\ket{2j+1}_b+\ket{2j+1}_a\ket{0}_b}{\sqrt{2}}\right)\Big\}
	\end{aligned}\label{ph-2}
\end{equation}
\end{widetext}
One can verify eq.(\ref{ph-2}) by the summation of density matrices $\ket{a}\bra{a}+\ket{b}\bra{b}=\frac{1}{2}(\ket{a}+\ket{b})(\bra{a}+\bra{b})+\frac{1}{2}(\ket{a}-\ket{b})(\bra{a}-\bra{b})$.

The estimation of eq.(\ref{ph-2}) is still hard, because we only know the click rate of the states $\ket{0}_a\ket{j}_b+\ket{j}_a\ket{0}_b$ with decoy states. We can use the Cauchy-Schwarz inequality shown in eq.(\ref{cauchy}) to estimate the upper bound of the phase error probability in eq.(\ref{2ph}).
\begin{widetext}
\begin{equation}
	\begin{aligned}
	\Tr(M^R(\sum_{j}\ket{a_j})(\sum_j\bra{a_j}))=\sum_j\Tr(M^R\ket{a_j}\bra{a_j})+\sum_{j\ne k}\Tr(M^R\ket{a_j}\bra{a_k})\\
	\le \sum_j\Tr(M^R\ket{a_j}\bra{a_j})+\sum_{j\ne k}\sqrt{\Tr(M^R\ket{a_j}\bra{a_j})\Tr(M^R\ket{a_k}\bra{a_k})}
	\end{aligned}\label{cauchy}
\end{equation}
\begin{equation}
	\begin{aligned}
	P_{ph}^{R-2}\le& p(1-p)\Bigg\{ \sum_{j=0}^\infty\frac{\text{e}^{-\mu}\mu^j}{j!}P^R(\frac{\ket{0}_a\ket{j}_b+\ket{j}_a\ket{0}_b}{\sqrt{2}}) \\
	&+\sum_{j\ne k}\text{e}^{-\mu}\sqrt{\frac{\mu^{2j+2k}}{2j!2k!}}\sqrt{P^R(\frac{\ket{0}_a\ket{2j}_b+\ket{2j}_a\ket{0}_b}{\sqrt{2}})}\sqrt{P^R(\frac{\ket{0}_a\ket{2k}_b+\ket{2k}_a\ket{0}_b}{\sqrt{2}})}\\
	&+\sum_{j\ne k}\text{e}^{-\mu}\sqrt{\frac{\mu^{2j+2k+2}}{(2j+1)!(2k+1)!}}\sqrt{P^R(\frac{\ket{0}_a\ket{2j+1}_b+\ket{2j+1}_a\ket{0}_b}{\sqrt{2}})}\sqrt{P^R(\frac{\ket{0}_a\ket{2k+1}_b+\ket{2k+1}_a\ket{0}_b}{\sqrt{2}})}\Bigg\}
	\end{aligned}\label{2ph}
\end{equation}
\end{widetext}
Comparing with eq.(\ref{pph}), the first line of eq.(\ref{2ph}) is the same when $\delta=0$, but some cross terms are added. With a larger phase error, the transmission distance will be decreased. 

The security key rate of this case is shown as 
\begin{equation}
	R^R=\frac{1}{2}(P_c^{R}(1-H_2(\frac{P_{ph}^{R-2}}{P_c^{R}}))-fP_{t-2}^RH_2(e_{bit-2})).
\end{equation}
Here $P_c^R$ is the same as the one in eq.(\ref{rate}), because the click rates of $\ket{\alpha}_a\ket{0}_b$ and $\ket{0}_a\ket{\alpha}_b$ are not influenced by the phase. $P_{t-2}^R$ is the right-click rate of sifted rounds and $e_{bit-2}$ is the bit error rate of the $M=2$ case.
\subsection{$M=4$}
With a same method, we can give the phase error probability of $M=4$ case, which is shown in eq.(\ref{4ph}).
\begin{widetext}
	\begin{equation}
	\begin{aligned}
	P_{ph}^{R-4}\le& p(1-p)\Bigg\{ \sum_{j=0}^\infty\frac{\text{e}^{-\mu}\mu^j}{j!}P^R(\frac{\ket{0}_a\ket{j}_b+\ket{j}_a\ket{0}_b}{\sqrt{2}}) \\
	&+\sum_{j\ne k}\text{e}^{-\mu}\sqrt{\frac{\mu^{4j+4k}}{4j!4k!}}\sqrt{P^R(\frac{\ket{0}_a\ket{4j}_b+\ket{4j}_a\ket{0}_b}{\sqrt{2}})}\sqrt{P^R(\frac{\ket{0}_a\ket{4k}_b+\ket{4k}_a\ket{0}_b}{\sqrt{2}})}\\
	&+\sum_{j\ne k}\text{e}^{-\mu}\sqrt{\frac{\mu^{4j+4k+2}}{(4j+1)!(4k+1)!}}\sqrt{P^R(\frac{\ket{0}_a\ket{4j+1}_b+\ket{4j+1}_a\ket{0}_b}{\sqrt{2}})}\sqrt{P^R(\frac{\ket{0}_a\ket{4k+1}_b+\ket{4k+1}_a\ket{0}_b}{\sqrt{2}})}\\
	&+\sum_{j\ne k}\text{e}^{-\mu}\sqrt{\frac{\mu^{4j+4k+4}}{(4j+2)!(4k+2)!}}\sqrt{P^R(\frac{\ket{0}_a\ket{4j+2}_b+\ket{4j+2}_a\ket{0}_b}{\sqrt{2}})}\sqrt{P^R(\frac{\ket{0}_a\ket{4k+2}_b+\ket{4k+2}_a\ket{0}_b}{\sqrt{2}})}\\
	&+\sum_{j\ne k}\text{e}^{-\mu}\sqrt{\frac{\mu^{4j+4k+6}}{(4j+3)!(4k+3)!}}\sqrt{P^R(\frac{\ket{0}_a\ket{4j+3}_b+\ket{4j+3}_a\ket{0}_b}{\sqrt{2}})}\sqrt{P^R(\frac{\ket{0}_a\ket{4k+3}_b+\ket{4k+3}_a\ket{0}_b}{\sqrt{2}})}\Bigg\}
\end{aligned}\label{4ph}
	\end{equation}
\end{widetext}

In this case, the phase error is close to the case of continuous phase randomization when $\delta=0$, because the phase errors mainly come from the items of 0, 1  and 2 photons and the cross terms of more than 4 photons have little influence.

The key rate of this case is shown as 
\begin{equation}
	R^R=\frac{1}{4}(P_c^{R}(1-H_2(\frac{P_{ph}^{R-4}}{P_c^{R}}))-fP_{t-4}^RH_2(e_{bit-4})).
\end{equation}
$P_{t-4}^R$ is the right-click rate of sifted rounds and $e_{bit-4}$ is the bit error rate.
\bibliography{refe.bib}
\end{document}